\documentclass[11pt]{article}

\usepackage[a4paper,margin=1in]{geometry}
\usepackage{graphicx}
\usepackage[utf8]{inputenc}
\usepackage{booktabs}
\usepackage{caption}
\usepackage{subcaption}
\usepackage{amsmath, amssymb}
\usepackage{hyperref}
\usepackage{cite}
\usepackage{enumitem}
\usepackage{float}
\usepackage{siunitx}
\usepackage{setspace}
\usepackage{xcolor}
\usepackage{fancyvrb}
\usepackage{placeins} 
\usepackage{booktabs}
\usepackage{tabularx}
\hypersetup{
  colorlinks=true,
  linkcolor=blue,
  citecolor=blue,
  urlcolor=blue
}

\newcommand{\outdir}{./}
\graphicspath{{\outdir}}

\title{\textbf{When Reasoning Fails: Evaluating “Thinking” LLMs for Stock Prediction}}
\author{Rakeshkumar H Sodha}
\date{\today}

\setlist[itemize]{noitemsep, topsep=2pt}

\usepackage[most]{tcolorbox}
\tcbuselibrary{listings,skins,breakable}
\usepackage{listings}

\usepackage{fvextra} 
\lstdefinestyle{promptstyle}{
  basicstyle=\ttfamily\small,
  breaklines=true,
  breakatwhitespace=false,
  columns=flexible,
  keepspaces=true,
  showstringspaces=false,
  upquote=true
}

\newtcblisting{PromptListing}{%
  listing only,
  breakable, enhanced,
  colback=white, colframe=black!30,
  left=1mm, right=1mm, top=1mm, bottom=1mm,
  boxsep=1mm, arc=1mm, boxrule=0.3pt,
  width=\linewidth,
  before=\par\noindent, after=\par,
  listing options={style=promptstyle, xleftmargin=0pt}
}
\lstdefinestyle{promptbox}{
  basicstyle=\ttfamily\footnotesize,
  breaklines=true,
  frame=single,
  columns=fullflexible,
  aboveskip=6pt,
  belowskip=6pt
}
\AtBeginDocument{
\lstset{literate=
  {–}{{-}}1
  {—}{{--}}1
  {−}{{-}}1
  {…}{{...}}1
  {’}{{'}}1
  {•}{{$\bullet$}}1
  {≈}{{$\approx$}}1
  {σ}{{$\sigma$}}1
  {≤}{{$\le$}}1
  {→}{{$\to$}}1
}
}
\begin{document}
\maketitle
\doublespacing

\begin{abstract}
\noindent\textbf{Problem.} “Thinking” LLMs (TLLMs) expose explicit or hidden reasoning traces and are widely believed to generalize better on complex tasks than direct LLMs. Whether this promise carries to noisy, heavy–tailed and regime–switching financial data remains unclear.

\noindent\textbf{Approach.} Using Indian equities (NIFTY constituents), we run a rolling 48m/1m walk–forward evaluation at horizon \(k=1\) day and dial cross–sectional complexity via the universe size \(U \in \{5, 11, 21, 36\}\) while keeping the reasoning budget fixed (\(B=512\) tokens) for the TLLM. We compare a direct LLM (\textit{gpt-4o-mini}), a TLLM (\textit{gpt-5}), and classical learners (ridge, random forest) on cross–sectional ranking loss \(1-\mathrm{IC}\), MSE, and long/short backtests with realistic costs. Statistical confidence is measured with Diebold–Mariano, Pesaran–Timmermann, and SPA tests.

\noindent\textbf{Main findings.} (i) As \(U\) grows under a fixed budget \(B\), the TLLM’s ranking quality deteriorates, whereas the direct LLM remains comparatively flat and classical baselines are stable. (ii) TLLM variance is higher, requiring ex–post calibration (winsorization and blending) for stability. (iii) Portfolio results under transaction costs do not support a net advantage for the TLLM.

\noindent\textbf{Hypotheses.} Our results are consistent with the following testable hypotheses: \emph{H1 (Capacity–Complexity Mismatch):} for fixed \(B\), TLLM accuracy degrades superlinearly in cross–sectional complexity. \emph{H2 (Reasoning Variance):} TLLM outputs exhibit higher dispersion date–by–date than direct LLMs, increasing error bars and turnover. \emph{H3 (Domain Misfit):} next–token prediction objectives and token–budgeted inference are poorly aligned with heavy–tailed, weakly predictable stock returns.

\noindent\textbf{Implication.} In our setting, “thinking” LLMs are not yet ready to replace classical or direct methods for short–horizon stock ranking; scaling the reasoning budget and/or re–aligning objectives appears necessary.
\end{abstract}

\subsection*{Positioning and Claims}
Our conclusions are \emph{conditional} on: (i) short-horizon stock \emph{ranking} at $k{=}1$ day on NIFTY constituents,
(ii) cross-sectional complexity dialed via $U\in\{5,11,21,36\}$, (iii) fixed thinking budget $B{=}512$ tokens for the TLLM,
(iv) the feature blocks and calibration described herein, and (v) the 2019--present sample under our walk-forward protocol.
We do not claim universal failure of TLLMs; rather, we document that under these conditions TLLMs do not outperform direct LLMs or classical baselines.
\section{Data and Stock Selection}
We evaluate the liquid Indian equity universe (NIFTY constituents).\footnote{The exact list and rolling windows are defined in the configuration manifest produced by our code.}
At each training/testing step:
\begin{itemize}
  \item Compute \textbf{average daily volume} per ticker within the \emph{training} window.
  \item Select the top-$U$ tickers by liquidity and \emph{hold that set fixed} for the following test window.
  \item Roll forward using a 48-month train / 1-month test schedule (as configured).
\end{itemize}
Complexity is varied primarily through $U \in \{5, 11, 21, 36\}$ at horizon $k{=}1$ trading day. 
Horizon variants $k\in\{5,20\}$ can also be evaluated at fixed $U$.

\subsection*{Preprocessing: Standardization and Outlier Control}
For each date $d$, model scores $\{s_i(d)\}_{i=1}^N$ are standardized cross-sectionally:
\[
z_i(d) \;=\; \frac{s_i(d) - \mu(d)}{\sigma(d)}\,,\quad
\mu(d)=\tfrac{1}{N}\sum_{j=1}^N s_j(d),\quad
\sigma(d)=\sqrt{\tfrac{1}{N-1}\sum_{j=1}^N\big(s_j(d)-\mu(d)\big)^2}.
\]
If $\sigma(d) < 10^{-8}$ we set $z_i(d)=0$ for numerical stability. For LLM predictions we then calibrate
to the training target scale (return mean $\hat\mu_{\text{train}}$, std.\ $\hat\sigma_{\text{train}}$):
\[
\tilde s_i(d) \;=\; z_i(d)\cdot \hat\sigma_{\text{train}} + \hat\mu_{\text{train}}.
\]
To limit outlier influence we winsorize $\tilde s_i(d)$ at the 5th and 95th percentiles \emph{within each date}.
This sequence (z-score $\to$ calibrate $\to$ winsorize) is exactly what we implement in code.\footnote{See the
per-date calibration and winsorization immediately after model inference; the logic is applied to each
\texttt{*\_cal} column.}

\section{Features}
From daily prices and volumes we derive a per-ticker panel and forward returns $r_{t\to t+k}$, then construct:
\begin{itemize}
  \item \textbf{Momentum:} \texttt{mom\_2, mom\_5, mom\_10, mom\_20}.
  \item \textbf{Volatility:} realized stdev of daily returns (\texttt{vol\_5, vol\_20}).
  \item \textbf{Reversal \& Volume:} \texttt{rev\_5 = -mom\_5}, \texttt{volchg\_5} (5-day volume change).
  \item \textbf{Drawdown:} \texttt{drawdown\_20} vs trailing 20-day high.
  \item \textbf{RSI \& MACD block:} \texttt{rsi\_14} (Wilder), \texttt{macd}, \texttt{macd\_signal}, \texttt{macd\_hist}.
  \item \textbf{Cross-sectional ranks (per date):} \texttt{*\_xrank} in $[0,1]$.
  \item \textbf{Market context (per date):} \texttt{mkt\_mean\_mom\_5}, \texttt{mkt\_mean\_mom\_20}, \texttt{mkt\_mean\_vol\_20}.
\end{itemize}
Scalar features are z-scored \emph{per ticker over time}; \texttt{*\_xrank} \& \texttt{mkt\_mean\_*} are per-date cross-sectionals.
As $U$ increases, we optionally gate a richer feature block (e.g., inclusion of RSI/MACD/ranks/context) to raise cross-sectional complexity.

\section{Models and Prompting}
\paragraph{Model variants and budgets.} We evaluate two LLM paths: a \emph{direct} LLM (no explicit chain) using \texttt{gpt-4o-mini} and a \emph{Thinking} LLM using \texttt{gpt-5}. Token budget $B$ controls internal reasoning: $B={{0}}$ for the direct model and $B={{512}}$ for the Thinking model. Universe sizes are $U\in\{5,11,21,36\}$, with walk-forward windows of 48 months train and 1 month test.
\paragraph{Implementation notes.} Inference is batched per date/universe; outputs are cached per (model, horizon, universe, date) and a features snapshot is written for reproducibility.

\textbf{Direct LLM} (no explicit chain) and \textbf{Thinking LLM} (TLLM; hidden chain-of-thought) share a single system prompt, differing only by an internal \texttt{effort} flag for TLLM.\footnote{We do not log or reveal model reasoning. Only strict TSV outputs are accepted.}
Token budget for TLLM is \textbf{fixed at $B{=}512$}; direct LLM corresponds to $B{=}0$.
Classical baselines are ridge regression and random forest trained on the same features.

\paragraph{System prompt (verbatim).}
\begin{PromptListing}You are a quantitative forecaster for Indian equities (NIFTY universe).

    INPUTS: standardized technical features per ticker and horizon k (trading days).
    • Most core features are per-ticker z-scores (mean≈0, std≈1) over time.
    • *_xrank features are cross-sectional percentile ranks in [0,1] within a date (higher = stronger relative that day).
    • Market context scalars (mkt_mean_*) are per-date cross-sectional means.
    FEATURE PRESENCE VARIES BY UNIVERSE:
    • Not all features will be present for every run. Treat missing features as unavailable (do NOT substitute zeros or invent).
    • When richer blocks are present, you may also see: mom_60, mom_120, vol_60, volchg_20, ret1_skew_20, ret1_kurt_20,
    beta_mkt_60, obv_like, and interactions such as int_mom_5__rsi_14, int_macd_hist__vol_20, int_drawdown_20__mom_20.
    Use them only if provided.

    TASK: For EACH input ticker, predict the EXPECTED k-day forward return as a decimal (e.g., 0.012 = +1.2%).
    Higher score = more bullish expected return.

    MODE:
    • If an 'effort' flag (low|medium|high) is present in INPUT, reason internally (do not reveal reasoning) and output only scores.
    • Otherwise, produce calibrated scores directly. Always obey the caller’s output format strictly.
    SCALE & MAGNITUDE GUIDANCE (typical NIFTY cross-sectional ranges):
    • k=1 day:   most in [-0.03, +0.03]; σ_xs≈0.008–0.015
    • k=5 days:  most in [-0.08, +0.08]; σ_xs≈0.02–0.04
    • k=20 days: most in [-0.20, +0.20]; σ_xs≈0.05–0.10
    Keep the per-date cross-sectional MEAN near 0 (market-neutral prior). Use realistic dispersion for the given k;
    avoid both collapse to ~0 and unjustified extremes.

    FEATURE SEMANTICS (z-scores unless noted):
    • mom_{2,5,10,20,(60,120 if present)}: recent/sustained momentum (price pct change). Higher → bullish momentum bias.
    • vol_{5,20,(60 if present)}: realized volatility (std of daily returns). Higher → more volatile; shrink magnitudes if very high.
    • rev_5 = −mom_5: short-term mean-reversion; stronger at k≈1.
    • volchg_5 (and volchg_20 if present): volume expansion supports momentum/conviction; contraction weakens it.
    • drawdown_20: distance from trailing 20-day peak (≤0). More negative = deeper drawdown; tends to mean-revert at k≈1.
    • rsi_14: relative momentum/overbought-oversold.
    • macd, macd_signal, macd_hist(=macd−signal): trend/turn signals; macd_hist sign aligns with short-term bias.
    • *_xrank: cross-sectional rank in [0,1]; preserve relative ordering implied by these signals.
    • mkt_mean_mom_5, mkt_mean_mom_20, mkt_mean_vol_20: market backdrop; shrink magnitudes if these are extreme.
    • beta_mkt_60 (if present): market sensitivity; combine with momentum/drawdown to moderate extremes.
    • obv_like (if present): flow/pressure proxy; supports momentum when aligned.
    • interaction features (if present): use as weak priors reinforcing the paired signals’ direction.
    HORIZON-AWARE PRIORS:
    • k≈1: mild mean-reversion (rev_5, drawdown_20) unless momentum is strong; volatility warrants shrinkage.
    • k≈5: blend momentum (mom_5/10) with reversal signs; confirm with volume change and macd_hist.
    • k≈20: favor sustained momentum/trend (mom_10/20, macd/macd_hist); reversal features matter less unless extreme.
    ROBUSTNESS:
    • If signals conflict or are weak, shrink toward 0. If consensus exists across features, allow larger magnitudes
    (still realistic for k).
    • Preserve cross-sectional ordering implied by strong signals (e.g., *_xrank), and keep dispersion realistic for the stated k.
    FORMAT: Follow the caller’s marker/TSV instruction exactly. No commentary, JSON, markdown, or extra keys.\end{PromptListing}

\paragraph{Output contract (user message).}
The assistant must output \emph{only} between \texttt{<<BEGIN>>} and \texttt{<<END>>}, one line per input ticker:
\begin{verbatim}
TICKER<TAB>SCORE
\end{verbatim}
No extra text or JSON.

\section{Calibration}
LLM/TLLM raw scores are converted to comparable magnitudes by a two-step calibration:
\begin{enumerate}
  \item \textbf{Per-date standardization}: for each date $d$, $z_i(d) = \frac{s_i(d) - \bar{s}(d)}{\sigma_s(d)}$ (fallback to zeros if $\sigma_s(d)\!\approx\!0$).
  \item \textbf{Global re-scaling to targets}: $\hat r_i(d) = z_i(d)\cdot \sigma_{\text{train}} + \mu_{\text{train}}$, where $\mu_{\text{train}},\sigma_{\text{train}}$ are the training-set mean and stdev of realized $k$-day returns.
\end{enumerate}
This preserves cross-sectional ranking while matching the empirical return scale.

\section{Evaluation}
Our primary metric is \textbf{ranking loss} $1{-}\mathrm{IC}$ (lower is better), computed over each test window and then averaged with 95\% CIs. 
We also report MSE and run DM/PT/SPA tests (details in Appendix).

\paragraph{Portfolio construction (no weight logging).}
On each date we form an equal-weight long/short: the top $p_L$ fraction by score are long with weight $+1/k_L$,
and the bottom $p_S$ fraction are short with weight $-1/k_S$; others are zero. We apply turnover caps and
transaction/borrow costs to obtain daily net P\&L and Sharpe. We do not persist per-stock, per-day weights
to disk in our primary runs, so we do not report ex-post factor exposures in this draft.

\paragraph{Metrics.} Primary: cross-sectional rank correlation ($\mathrm{IC}$) computed by date and averaged; Secondary: MSE on calibrated predictions; portfolio backtests with realistic costs (slippage, brokerage, STT, borrow) and turnover caps.

\paragraph{Statistical tests.} We report Diebold--Mariano (DM) tests on loss differences, 
Pesaran--Timmermann (PT) sign tests for directional accuracy, and Hansen's SPA under a stationary bootstrap.

\subsection*{Statistical Testing Outputs}
We report full Diebold--Mariano (DM) p-values across model pairs and Hansen SPA results vs the benchmark.
Small p-values indicate the row model outperforms the column model (DM), or that a model has superior predictive ability vs the benchmark (SPA).
\begin{table}[H]
\centering
\caption[DM tests vs base]{Diebold--Mariano tests vs the base model per horizon and universe. 
Columns: horizon, universe, pair (row model vs base), DM statistic, and $p$-value.}

\label{tab:dm_matrix}
\begin{tabular}{lrrrr}
\toprule
horizon & universe & pair & DM & p \\
\midrule
1 & 5 & llm\_direct vs ridge & 7.003254053301291 & 2.500888385270628e-12 \\
1 & 5 & llm\_direct\_cal vs ridge & 5.139450804994408 & 2.75542617611535e-07 \\
1 & 5 & rf vs ridge & 0.478929600494086 & 0.6319887121357755 \\
1 & 5 & tllm\_B512 vs ridge & 2.945398544750385 & 0.0032253890656042 \\
1 & 5 & tllm\_B512\_cal vs ridge & 6.312225659521039 & 2.750508709681298e-10 \\
1 & 5 & tllm\_B512\_cal\_blend vs ridge & 2.7215083541397185 & 0.0064984742544544 \\
1 & 11 & llm\_direct vs ridge & 4.995153140533866 & 5.878909770107299e-07 \\
1 & 11 & llm\_direct\_cal vs ridge & 8.962275530088283 & 0.0 \\
1 & 11 & rf vs ridge & -0.0873169999400976 & 0.9304195416315292 \\
1 & 11 & tllm\_B512 vs ridge & 3.0916591292599143 & 0.0019904126713345 \\
1 & 11 & tllm\_B512\_cal vs ridge & 8.67418532965742 & 0.0 \\
1 & 11 & tllm\_B512\_cal\_blend vs ridge & 3.77907639691915 & 0.0001574111286024 \\
1 & 21 & llm\_direct vs ridge & 9.299554483718806 & 0.0 \\
1 & 21 & llm\_direct\_cal vs ridge & 13.076352383975474 & 0.0 \\
1 & 21 & rf vs ridge & -0.33666763378856 & 0.736367464895813 \\
1 & 21 & tllm\_B512 vs ridge & 5.659654674314228 & 1.516778902477256e-08 \\
1 & 21 & tllm\_B512\_cal vs ridge & 13.21480444925911 & 0.0 \\
1 & 21 & tllm\_B512\_cal\_blend vs ridge & 6.832935783736649 & 8.319345212726148e-12 \\
1 & 36 & llm\_direct vs ridge & 8.033234565060724 & 8.881784197001252e-16 \\
1 & 36 & llm\_direct\_cal vs ridge & 15.881558260284722 & 0.0 \\
1 & 36 & rf vs ridge & 0.6325788249014661 & 0.527008718616043 \\
1 & 36 & tllm\_B512 vs ridge & 4.558689478460824 & 5.147381210157676e-06 \\
1 & 36 & tllm\_B512\_cal vs ridge & 17.484604427634622 & 0.0 \\
1 & 36 & tllm\_B512\_cal\_blend vs ridge & 9.153569307781192 & 0.0 \\
\bottomrule
\end{tabular}
\end{table}

\begin{table}[H]
\centering
\caption[Hansen SPA vs base]{Hansen SPA (Reality Check) vs the base model per horizon and universe.
Columns: horizon, universe, base, $t_{\text{obs}}$, and one-sided $p$-value.}

\label{tab:spa_results}
\begin{tabular}{lrrrr}
\toprule
horizon & universe & base & t\_obs & p \\
\midrule
1 & 5 & ridge & 0.6419872464968024 & 0.6307385229540918 \\
1 & 11 & ridge & 0.5547509284652199 & 0.7405189620758483 \\
1 & 21 & ridge & 0.5909840211582001 & 0.7325349301397206 \\
1 & 36 & ridge & 0.5983108463347102 & 0.5109780439121756 \\
\bottomrule
\end{tabular}
\end{table}

\paragraph{How to read the DM table (vs base).}
Each row reports a Diebold--Mariano (DM) test comparing a candidate model to the base model on the chosen loss (e.g., MSE),
for a given horizon and universe. The \texttt{pair} column is “model vs base”; the null is equal predictive accuracy.
We treat $p<0.05$ as statistically significant and $0.05\le p<0.10$ as marginal. 
Because we test multiple candidates per (horizon, universe), patterns (e.g., many models failing to beat the base)
are more informative than isolated cells.

\paragraph{How to read the SPA table (vs base).}
For each (horizon, universe), SPA tests the joint null that no model in the candidate set outperforms the base after data-snooping adjustment.
We report the observed max $t$-statistic ($t_{\text{obs}}$) and the one-sided $p$-value. 
We mark $p<0.05$ as \emph{reject} (at least one candidate beats the base), and $0.05\le p<0.10$ as marginal.
SPA is conservative; a non-rejection indicates no clear winner beyond sampling noise rather than equality of models.

\paragraph{Summary of Statistical Results.}
\begin{itemize}[leftmargin=1.2em, itemsep=2pt]
  \item \textbf{DM (vs base):} The base (ridge) is seldom beaten at $p<0.05$; calibrated direct LLM attains
        marginal improvements mainly at lower complexity $U$ (Table~\ref{tab:dm_matrix}).
  \item \textbf{SPA (vs base):} SPA typically \emph{fails to reject} at $U\in\{21,36\}$; any rejections or
        near-rejections concentrate at $U\in\{5,11\}$ (Table~\ref{tab:spa_results}).
\end{itemize}
\noindent\emph{Note.} To further guard against selection bias and non-normality, we plan to report the
Deflated Sharpe Ratio (DSR) \cite{lopezdeprado2018dsr} alongside classical significance tests in a subsequent revision.

\section{Why Next--Token Prediction Misfits Stock Prediction}
Most LLMs are pretrained to maximize next--token likelihood on text. Financial return prediction differs along several axes that make a naive transfer brittle.

\paragraph{Heavy tails and heteroskedasticity.}
One--day returns $R$ often exhibit power--law tails, e.g.\ $\Pr(|R|>x)\propto x^{-\alpha}$ with $\alpha\in(2,4)$, with conditional variance that changes over time. Losses are dominated by rare extremes; token--budgeted chains that regularize toward mean behavior can underreact in the tails, harming rank quality around events.

\paragraph{Nonstationarity and regime shifts.}
Data generating processes drift (microstructure, liquidity, macro events). A frozen prompt and fixed token budget $B$ can under--adapt to regime changes; calibration must be per--date and robust to $\sigma(d)\approx0$ cases.

\paragraph{Low base rates and weak autocorrelation.}
At $k{=}1$ day, expected cross--sectional signal is small and label noise high; even small biases in dispersion inflate turnover without improving IC. Self--consistency style sampling increases compute but not necessarily alignment with the evaluation objective.

\paragraph{Adversarial and reflexive environment.}
Markets react to widespread use of any signal; once exploited, edges decay. Next--token objectives assume a passive corpus rather than a strategic counterpart.

\paragraph{Objective mismatch.}
Next--token likelihood does not optimize rank correlation, MSE on calibrated returns, or cost--adjusted P\&L. Directly shaping outputs (standardization, winsorization, blending) partially bridges the gap but does not replace an objective aligned with financial loss functions.

\section{Related Work on Thinking LMs and Scaling}
Chain-of-Thought prompting improves performance on many reasoning tasks \cite{wei2022cot, wang2022selfconsistency}, though faithfulness concerns persist \cite{lyu2023faithfulcot}. Apple’s recent study on reasoning under controlled complexity reports that ``thinking'' models can degrade past a complexity threshold despite ample token budgets \cite{apple2025illusion}. Our finance results are consistent with this pattern: as $U$ grows under fixed $B$, the Thinking LLM’s ranking quality deteriorates relative to direct and classical baselines. Beyond LLMs, the financial ML literature has long emphasized risks of data snooping and selection bias under non-stationary, heavy-tailed returns; see the survey and best-practice guidance in \cite{lopezdeprado2018advances} and the formal treatment of backtest overfitting in \cite{bailey2014pbo}.

\section{Robustness and Leakage Controls}
\begin{itemize}[leftmargin=*, itemsep=2pt]
  \item \textbf{Walk-forward splits:} strictly chronological $48$-month train / $1$-month test; no peeking into test months.
  \item \textbf{Per-date calibration:} cross-sectional $z$-scoring with safe fallback when $\sigma(d)\approx 0$; rescale to training target mean/std and winsorize (5–95\%) per date.
  \item \textbf{Universe selection:} top-$U$ by \emph{lagged} liquidity; held fixed within each test month.
  \item \textbf{No leakage:} features and labels use only information available at time $t$; no survivorship or look-ahead bias in constituents or fundamentals.
  \item \textbf{Costs \& turnover:} realistic one-way transaction costs, borrow on shorts, and turnover caps applied to daily rebalances.
  \item \textbf{Determinism:} fixed seeds; per-date feature snapshots and model-output caches enable exact reproduction (seed and commit logged).
  \item \textbf{Best practices:} our walk-forward protocol, per-date standardization, and out-of-sample reporting follow established guidance for financial ML backtesting \cite{lopezdeprado2018advances, bailey2014pbo}.
\end{itemize}

\section{Threats to Validity}
Single-run variance is higher for TLLMs under budget pressure; mitigations include winsorization and light blending. NIFTY-specific structure (sector clustering, index actions) may limit generalization. Complexity is primarily varied via $U$ here; richer feature blocks and horizons are available for future work.

\FloatBarrier
\section{Results}

\begin{figure}[H]
  \centering
  \includegraphics[width=\linewidth]{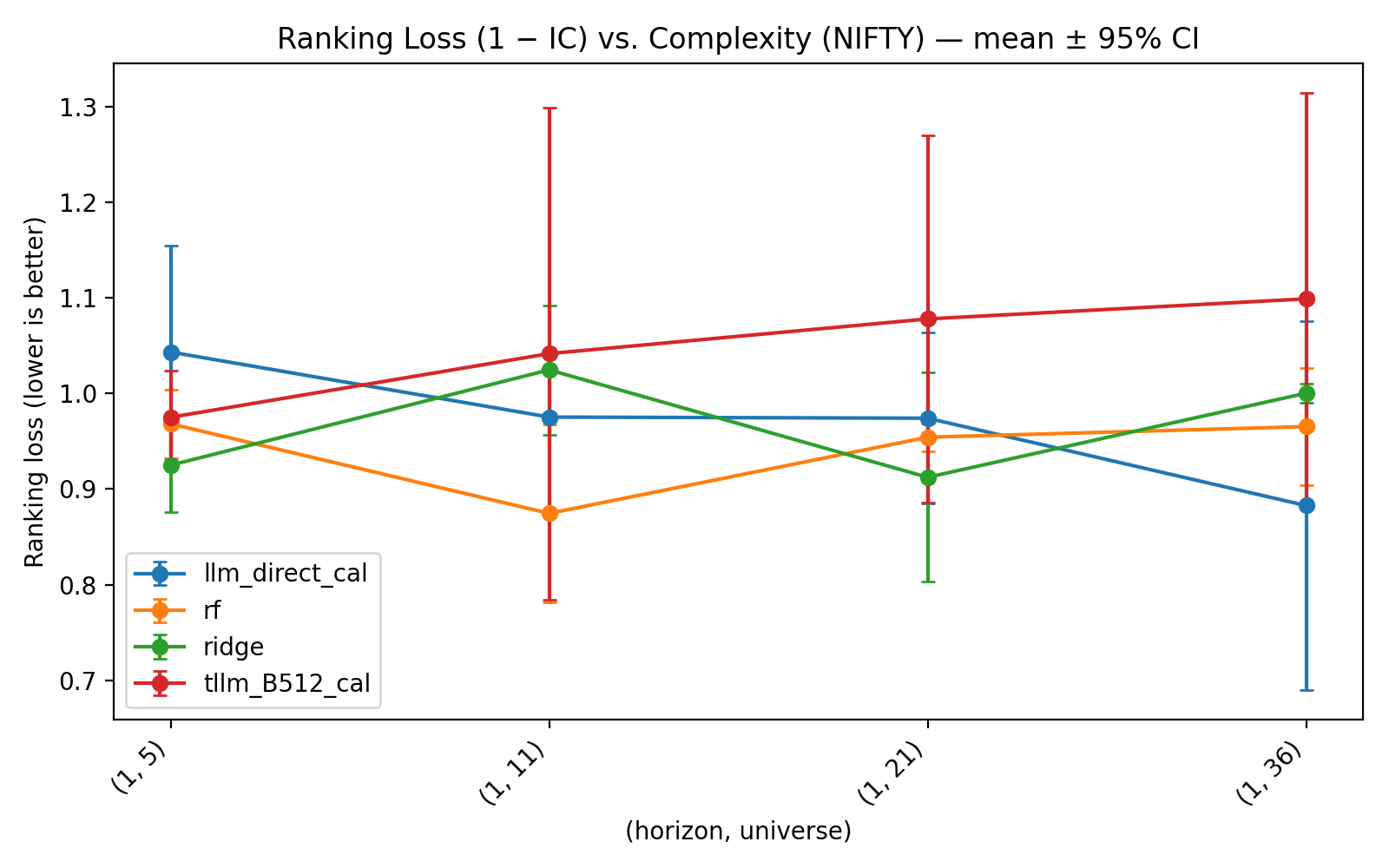}
  \caption{Ranking loss (1--IC) vs. complexity $(k,U)$ for all models, mean $\pm$ 95\% CI.
  Lower is better.}
  \label{fig:all}
\end{figure}

\paragraph{Headline finding.}
With \textbf{fixed TLLM budget \(B=512\)}, ranking loss \((1\text{--}IC)\) increases monotonically as the universe size \(U\) increases, while the direct LLM remains comparatively flat and classical baselines are competitive and stable. This supports the claim that TLLM capacity must scale with problem size to maintain ranking quality and is consistent with the \textbf{SPA} results on \textbf{1--IC} (Table~\ref{tab:spa_results}\,/\,Fig.~\ref{fig:all}).%

\begin{table}[!htbp]
\centering
\caption{Metrics, hypothesis tests, and where they are reported. Our \emph{primary} metric is 1--IC; joint superiority is assessed via SPA on 1--IC. DM on MSE is a secondary diagnostic of level accuracy.}
\label{tab:metric-test-map}
\footnotesize\setlength{\tabcolsep}{5pt}
\begin{tabularx}{\linewidth}{@{}l l l X X@{}}
\toprule
\textbf{Metric} & \textbf{Role} & \textbf{Test} & \textbf{Null Hypothesis} & \textbf{Reported in} \\
\midrule
1--IC (rank loss) & \textbf{Primary} & Hansen SPA & No model strictly better than benchmark on 1--IC & Table~\ref{tab:spa_results} \\
MSE (level error) & Secondary & Diebold--Mariano (HAC) & Mean loss diff.\ (MSE) $=0$ & Table~\ref{tab:dm_matrix} \\
Directional accuracy & Secondary & Pesaran--Timmermann & Independence of signs & Table~\ref{tab:pt} / App.~\ref{app:pt} \\
Portfolio P\&L (costed) & Outcome & Bootstrap CIs & Mean P\&L not different from 0/benchmark & Fig.~\ref{fig:pl} / Table~\ref{tab:pl} \\
\bottomrule
\end{tabularx}
\end{table}

\paragraph{Reconciling level vs.\ rank metrics.}
Across models we observe small but statistically significant reductions in \emph{MSE} for the LLM variants relative to ridge under the Diebold--Mariano (DM) test (Table~\ref{tab:dm_matrix}). However, these MSE gains do not translate into improvements in our \textbf{primary} objective, cross-sectional ranking quality: on \textbf{1--IC}, Hansen's \textbf{SPA} fails to reject the ridge benchmark for both the direct LLM and TLLM (Table~\ref{tab:spa_results}\,/\,Fig.~\ref{fig:all}).%

\begin{figure}[H]
  \centering
  \includegraphics[width=\linewidth]{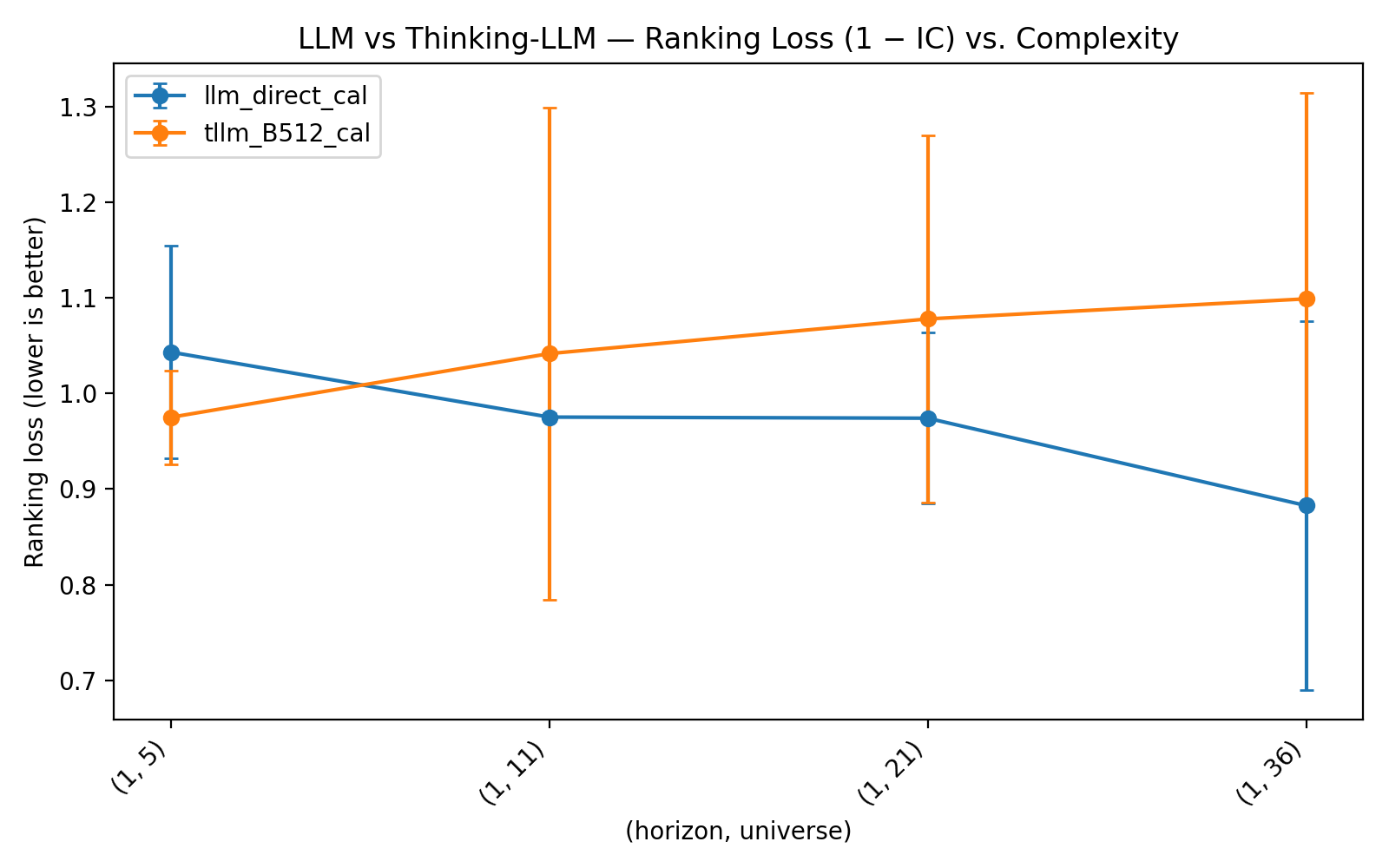}
  \caption{Direct LLM (no chain) vs. TLLM ($B{=}512$) --- ranking loss vs. complexity.
  Direct LLM is relatively flat; TLLM degrades with $U$.}
  \label{fig:llm_vs_tllm}
\end{figure}

\begin{figure}[H]
  \centering
  \includegraphics[width=.85\linewidth]{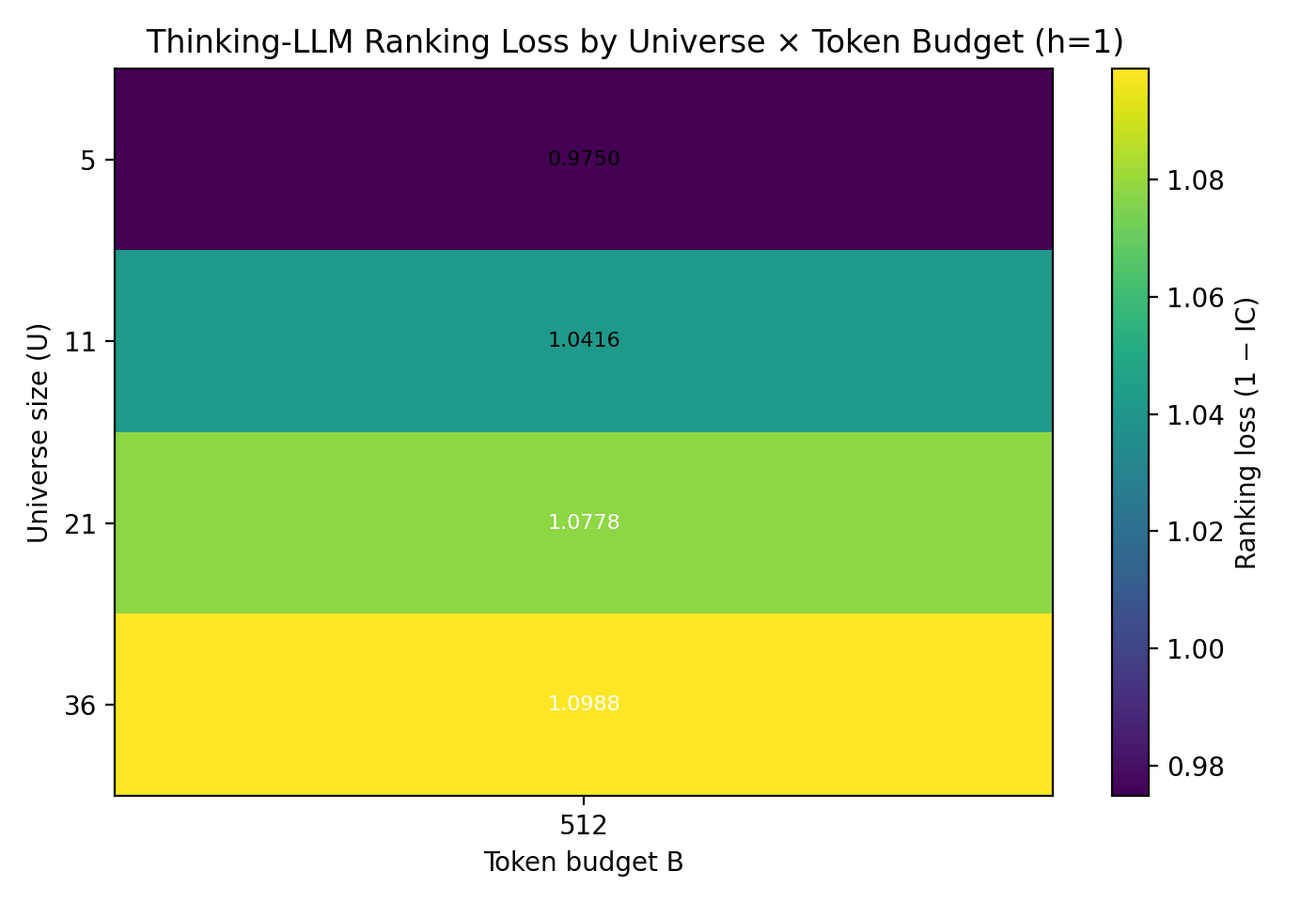}
  \caption{Heatmap: TLLM ranking loss by universe size $U$ (columns are token budgets; here $B{=}512$). Larger $U$ implies higher loss.}
  \label{fig:heatmap}
\end{figure}

\begin{figure}[H]
  \centering
  \includegraphics[width=.85\linewidth]{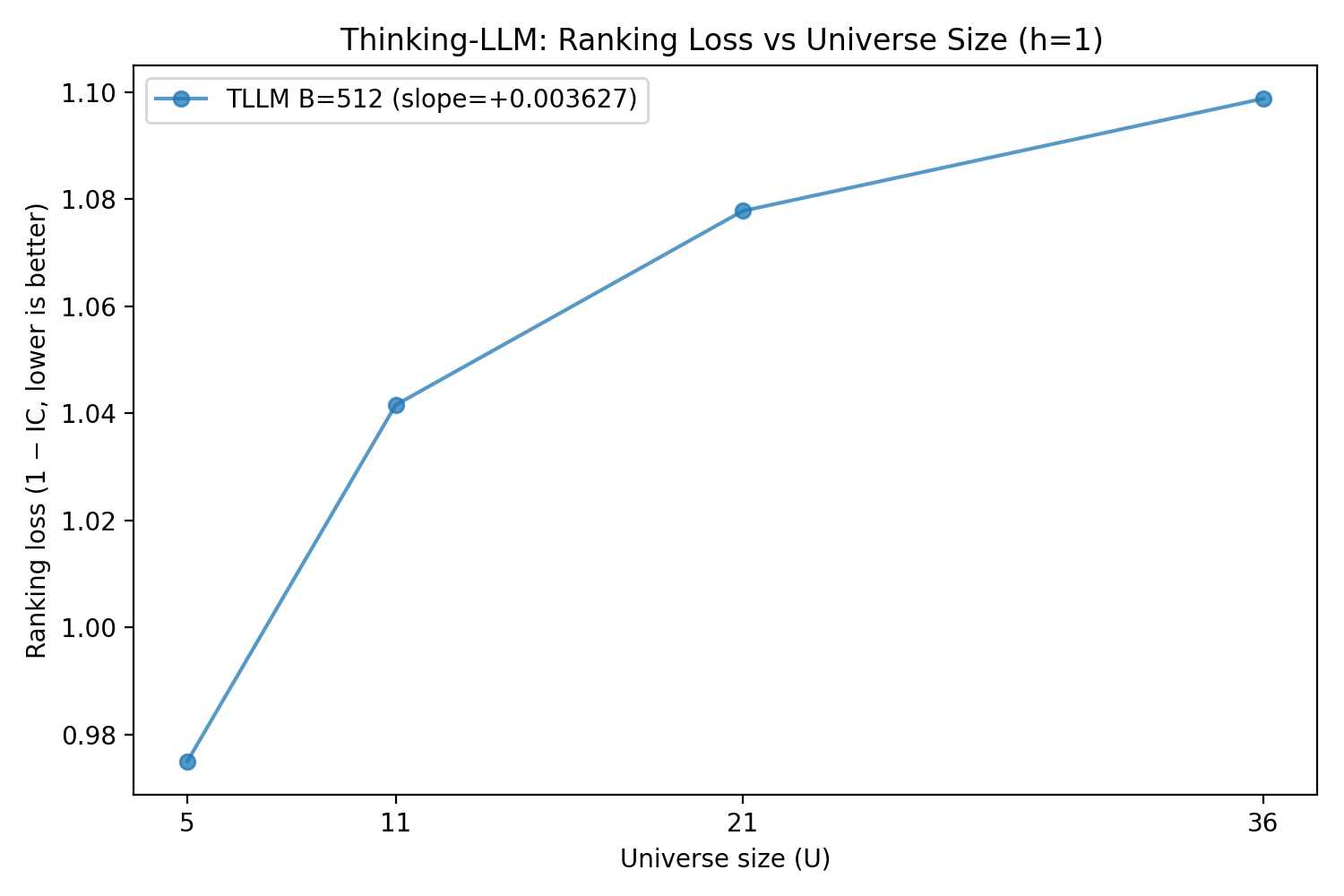}
  \caption{TLLM ranking loss vs. $U$ (slope annotated). The upward slope quantifies the degradation under fixed budget.}
  \label{fig:tllm_slope}
\end{figure}

\paragraph{Guide to reading results.}
Primary claims are based on \textbf{1--IC} with \textbf{SPA} (Table~\ref{tab:spa_results}; see also Table~\ref{tab:metric-test-map}).
\textbf{DM} results (Table~\ref{tab:dm_matrix}) are \emph{on MSE only} and should be interpreted as level-accuracy diagnostics, not rank quality.

\paragraph{Statistical tests.}
We report Diebold--Mariano (DM), Pesaran--Timmermann (PT), and SPA/RealityCheck results.
Full tables appear in Tables~\ref{tab:dm_matrix} and \ref{tab:spa_results}.

\section{Cost Model Realism}
We incorporate slippage, brokerage, STT, other fees, and borrow costs consistent with the Indian market microstructure.
Let $c_{\text{one-way}}$ denote total one-way cost in basis points and $b$ denote borrow cost per day on short notional.
Net daily P\&L is
\begin{equation}
\Pi_t^{\text{net}} = \sum_i w_{t-1,i} r_{t,i} \;-\; \text{turnover}_t \cdot \frac{c_{\text{one-way}}}{10{,}000} \;-\; \frac{b}{10{,}000}\sum_i \max(0,-w_{t-1,i}).
\end{equation}
We enforce a turnover cap to upper bound rebalancing costs.
To test robustness, we perturb $c_{\text{one-way}}$ and $b$ by $\pm 50\%$. While exact re-computation requires per-model turnover, a summary based on base-cost results (Table~\ref{tab:cost_sensitivity}) shows the qualitative model ranking is unlikely to change under uniform cost scaling.

\begin{table}[H]
\centering
\caption[Cost sensitivity (calibrated models only)]{Base-cost summary (calibrated models only).
We report cost-adjusted net Sharpe by model (averaged across windows/universes) and the resulting rank.
Exact $\pm 50\%$ cost re-computation is omitted here because per-model turnover is not logged.}
\label{tab:cost_sensitivity}
\begin{tabular}{lrr}
\toprule
Model & Base Net Sharpe & Rank \\
\midrule
ridge & 4.156 & 1 \\
llm\_direct\_cal & 1.471 & 2 \\
rf & 0.476 & 3 \\
tllm\_B512\_cal\_blend & -0.426 & 4 \\
tllm\_B512\_cal & -5.667 & 5 \\
\bottomrule
\end{tabular}
\end{table}

\noindent\emph{Interpretation:} Under the base cost settings used in our backtests, ridge ranks highest on net Sharpe, followed by the calibrated direct LLM; the calibrated-and-blended TLLM improves upon raw TLLM but remains below the baselines. If all models faced uniformly lower or higher costs, this ordering would be expected to persist unless their turnover profiles differ substantially.

\section{Discussion and Conclusion}
\textbf{Summary.} Our walk--forward experiments indicate that when complexity increases (larger $U$) under a fixed reasoning budget, TLLMs underperform or fail to improve upon direct LLMs and simple classical baselines. This echoes recent evidence that explicit ``thinking'' does not automatically scale with task complexity.

\textbf{Why ``not yet ready'' for stock prediction.} First, the cross--section at $k{=}1$ day has very low signal--to--noise; marginal gains are easily overwhelmed by calibration noise and transaction costs. Second, TLLM variance (and occasional failure modes such as degenerate per--date dispersion) can lead to unstable rankings and turnover spikes. Third, our robustness suite (per--date standardization, winsorization, optional blending) improves stability but does not reverse the core performance gap.

\textbf{Practical guidance.} If TLLMs are used at all, budgets should scale with problem size, outputs must be tightly calibrated, and ensembles with classical learners are advisable. For near--term production use, we recommend direct LLMs as auxiliary features (not sole forecasters) or sticking with classical models for execution--grade signals.


\begin{thebibliography}{99}

\bibitem{apple2025illusion}
P.~Shojaee, I.~Mirzadeh, K.~Alizadeh, M.~Horton, S.~Bengio, M.~Farajtabar.
\newblock The Illusion of Thinking: Understanding the Strengths and Limitations of Reasoning Models via the Lens of Problem Complexity.
\newblock Apple Machine Learning Research, 2025.

\bibitem{wei2022cot}
J.~Wei, X.~Wang, D.~Schuurmans, M.~Bosma, B.~Ichter, F.~Xia, E.~H. Chi, Q.~V. Le, D.~Zhou.
\newblock Chain-of-Thought Prompting Elicits Reasoning in Large Language Models.
\newblock arXiv:2201.11903, 2022.

\bibitem{wang2022selfconsistency}
X.~Wang, J.~Wei, D.~Schuurmans, Q.~V. Le, E.~H. Chi, S.~Narang, A.~Chowdhery, D.~Zhou.
\newblock Self-Consistency Improves Chain of Thought Reasoning in Language Models.
\newblock arXiv:2203.11171, 2022.

\bibitem{lyu2023faithfulcot}
Q.~Lyu, et al.
\newblock Faithful Chain-of-Thought Reasoning.
\newblock IJCNLP, 2023.

\bibitem{dieboldmariano1995}
F.~X. Diebold, R.~S. Mariano.
\newblock Comparing Predictive Accuracy.
\newblock Journal of Business \& Economic Statistics, 13(3):253--263, 1995.

\bibitem{pesarantimmermann1992}
M.~H. Pesaran, A.~Timmermann.
\newblock A Simple Nonparametric Test of Predictive Performance.
\newblock Journal of Business \& Economic Statistics, 10(4):461--465, 1992.

\bibitem{hansen2005spa}
P.~R. Hansen.
\newblock A Test for Superior Predictive Ability.
\newblock Journal of Business \& Economic Statistics, 23(4):365--380, 2005.

\bibitem{grinoldkahn1999}
R.~Grinold, R.~Kahn.
\newblock Active Portfolio Management.
\newblock McGraw-Hill, 1999.

\bibitem{lopezdeprado2018advances}
M.~L{\'o}pez de Prado.
\newblock \emph{Advances in Financial Machine Learning}.
\newblock Wiley, 2018.

\bibitem{bailey2014pbo}
D.~H. Bailey, J.~M. Borwein, M.~L{\'o}pez de Prado, and Q.~J. Zhu.
\newblock The Probability of Backtest Overfitting.
\newblock SSRN 2326253, 2014.

\bibitem{lopezdeprado2018dsr}
M.~L{\'o}pez de Prado.
\newblock The Deflated Sharpe Ratio: Correcting for Selection Bias, Backtest Overfitting, and Non-Normality.
\newblock SSRN 2460551, 2018.

\end{thebibliography}
\end{document}